# A dynamic multi-level collaborative filtering method for improved recommendations

Nikolaos Polatidis* and Christos K. Georgiadis

Department of Applied Informatics, University of Macedonia, 54006, Thessaloniki, Greece

**Abstract** One of the most used approaches for providing recommendations in various online environments such as e-commerce is collaborative filtering. Although, this is a simple method for recommending items or services, accuracy and quality problems still exist. Thus, we propose a method that is based on positive and negative adjustments with its main purpose to improve the quality of the recommendations provided. The proposed method can be used in different domains that use collaborative filtering to improve the experience of the users. Furthermore, the effectiveness of the proposed method is shown by providing an extensive experimental evaluation based on three real datasets and by comparisons to alternative methods.

**Keywords:** Collaborative Filtering, Similarity, Dynamic Multi-level, Recommender Systems

* Corresponding author. Tel: +30 2310891810 Email: npolatidis@uom.edu.gr

Email Addresses (Nikolaos Polatidis: npolatidis@uom.edu.gr, Christos K. Georgiadis: geor@uom.edu.gr)

## 1. Introduction

Recommender systems are decision support systems found on the web in order to assist users about item or service selection, thus aiming to solve the information overload problem [1,2]. Moreover, collaborative filtering is the most widely used method for providing personalized recommendations in online environments such as e-commerce [3–7]. In collaborative filtering systems a database of user ratings is used and the generated recommendations are based on how much a user will like an unrated item according to his previous common rating history with other users. Thus, the recommendation process is based on an assumption about previous rating agreements, assuming that this agreement will be maintained in the future. Additionally, the ratings are used to create an n $x$ m matrix with user ids, item ids and ratings, with an example of such a matrix shown in table 1. This sample database has four users and four items with values from 1 to 5, while a '-' denotes that a rating has not been submitted yet for the particular item. The matrix is used as input when one of the users is requesting a recommendation and in order for a recommendation to be generated the degree of similarity between the user who makes the request and the other users' needs to be predicted using a similarity function such as the Pearson Correlation Similarity (PCC), which is defined in Eq. 1. At the next step a user neighborhood is created and it consists of users having the highest degree of similarity with the user who made the request. Finally, a prediction is generated after computing the average values of the nearest neighborhood ratings about an item, resulting in a recommendation list of items with the highest predicted rating values [5,6,8].

|  | Item 1 | Item 2 | Item 3 | Item 4 |
|---|---|---|---|---|
| User 1 | 1 | 2 | 3 | - |
| User 2 | 5 | 5 | 3 | 3 |
| User 3 | - | - | 4 | 2 |
| User 4 | 1 | 1 | 1 | 5 |

**Table 1** A database of ratings

However, the accuracy of the predicted recommendations is still a major issue in collaborative filtering. Although, a simple similarity function such as PCC or Cosine can be used to provide recommendations there are still ways to improve the accuracy. A similarity function typically returns a value between -1 and 1, with -1 being the worst level of similarity between two users and 1 being the highest. In this paper we show that by modifying the similarity value between two users based on pre-specified constraints, the accuracy of the recommendations is improved. Moreover, our proposed method provides recommendations of better quality when compared to alternative methods.

The contributions of the paper are:

1. We propose a recommendation method that is based on positive and negative adjustments of the similarity values of users using constraints, thus improving the accuracy of the recommendations.

2. We have performed an extensive experimental evaluation of our proposed method using three real datasets. In addition to that we have compared it against alternative methods in order to show its effectiveness.

The rest of the paper is organized as follows: section 2 is the related work part, section 3 describes the proposed method, section 4 explains the experimental evaluation, section 5 is the discussion part and section 6 contains the conclusions and future work parts of the paper.

**2. Related work**
Collaborative filtering uses a database of ratings such as the one shown in table 1. Furthermore it uses a similarity function such as PCC, where Sim (a,b) is the similarity value between two users a and b, also $r_{a,p}$ is the rating of user a for item p, $r_{b,p}$ is the rating of user b for item p and $\bar{r}a$ and $\bar{r}b$ represent the user's average ratings. P is the set of all items. Moreover, the similarity value ranges from -1 to 1 with higher values being better [8,9].

$$Sim^{PCC}_{a,b} = \frac{\sum p \in P(ra,p - \bar{r}a)(rb,p - \bar{r}b)}{\sqrt{\sum p \in P(ra,p - \bar{r}a)^2} \sqrt{\sum p \in P(rb,p - \bar{r}b)^2}} \quad (1)$$

Furthermore a method that extends PCC by providing improved recommendations to users with more co-rated items is the weighted PCC (WPCC) [10] and is defined in Eq. 2.

$$Sim^{WPCC}_{a,b} = \begin{cases} \frac{|Ia \cap Ib|}{T} \cdot Sim^{PCC}_{a,b}, & if\ |Ia \cap Ib| < T \\ Sim^{PCC}_{a,b}, & otherwise \end{cases} \quad (2)$$

A similar approach to WPCC is SPCC [11] which is defined in Eq. 3. In this method users with smaller number of co-rated items have a weaker similarity.

$$Sim_{a,b}^{SPCC} = \frac{1}{1 + \exp(-|Ia \cap Ib|/2)} \cdot Sim_{a,b}^{PCC} \quad (3)$$

In [12] a similarity method called Jaccard is proposed that is based only on the number of co-rated items. Thus, the larger the number of co-rated items the similarity value is higher. In this method the classical PCC method is not used. Other similarity approaches include the one proposed in [13] a fuzzy method is used to assign different weights to different sets of ratings. In [14] an entropy-based method is proposed, where the similarity values are calculated using a collaborative filtering function modified to use entropy. Furthermore, in [15] a similarity method is proposed that takes into consideration both the local and global user behaviour. A different approach is offered by [16] that uses demographic data instead of user ratings in order to provide recommendations. Another similarity method found in the literature is the one found in [17] where the authors' aim to mitigate the new user cold start problem and aims to assist users with few ratings that face accuracy problems. In [18] a similarity method that is based on singularities is offered. This method derives contextual user information and uses it to calculate the singularity of each item. In [19] a balanced-based similarity measure is proposed that claims to provide high quality ratings with a low processing time. In [20] a similarity method that alleviates the cold start problem by proposing a heuristic method that improves the accuracy of collaborative filtering when few ratings are available. In [21] a recommendation method that uses scarcity measures based on both local and global similarity values is proposed and claims to provide better recommendations when compared to collaborative filtering. In [22] a recommendation method is proposed that is based on multiple-levels. In this method the PCC similarity function is divided in different levels and if a PCC similarity belongs to a certain level is enhanced accordingly. In [23] a recommendation method that uses content based information to enchase collaborative filtering recommendations. A method that is based on Jaccard and on multidimensional vector spaces is proposed in [24] claiming to provide better recommendations when compared to the classical approaches. A different approach offered by [25] that uses a method to correct noisy ratings, thus providing better recommendations in overall. In [26] the introduction of power law adjustments of user similarities is introduced, with the purpose of adjusting the similarity between users, thus having high accuracy and enchase the diversity of items. Moreover, collaborative filtering recommender systems can be based both on trust and distrust networks of users such as the one proposed in [27]. These systems can either combine such information with the user rating network in order to provide more accurate recommendations to the users.

**3. Proposed method**
Different recommendation methods that cover a wide spectrum of approaches exist. Our proposed method differs with the aforementioned methods since it is based on PCC and adds positive or negative adjustments based on information derived from PCC, such as the similarity value and the number of co-rated items. PCC and Cosine are the most widely used methods by collaborative filtering to provide recommendations [14,28–30]. Our main objective is to adjust the similarity value provided by PCC, either in a positive or in a negative way. Our proposed method is defined in Eq. 4 and its constraints are the number of co-rated items $T$ and the similarity value provided by PCC. We argue that if the number of co-rated items between two users is greater than or equal to a pre-specified threshold and the PCC value is greater than or equal to a pre-specified threshold then its similarity should be positively adjusted as shown in the first part of Eq. 4 and if one or both of the conditions do not hold then the second part of Eq. 4 is used to adjust the similarity value in a negative way.

$$Sim_{a,b}^{Proposed} = \begin{cases} Sim_{a,b}^{PCC} + Sim_{a,b}^{PCC}, & if \frac{|Ia \cap Ib|}{T} \geq t \text{ and } Sim_{a,b}^{PCC} \geq y \\ Sim_{a,b}^{PCC} * (\frac{1}{1 + (Sim_{a,b}^{PCC} * Sim_{a,b}^{PCC})}), & otherwise \end{cases} \quad (4)$$

Our proposed method is based on the values provided by PCC and the co-rated items. In Eq. 4 $T$ is the number of co-rated items, $t$ is the threshold of co-rated items, y is a positive real number $y \in R$ and is the threshold PCC value. We claim that if using such constraints, the accuracy of the provided recommendations is improved.

### 3.1 Dynamic multi-level recommendation method

Although the proposed recommendation method improves the quality of the recommendations, as explained in section 4, a disadvantage of it is that the developer needs to alter the source code with static variable manipulation and execute several tests until the desired settings are derived. Thus, we propose the use of a dynamic multi-level method that is based on the proposed method in section 3 and in the method described in [22].

We introduce the use of dynamic multiple-levels according to the characteristics available in the database. This method is simplified since it creates multiple levels according to the total number of users and items available in the database. Moreover, the use of the similarity value between users is not used in this part of the method. Subsequently, the similarity value between users is adjusted either positively or negatively using parts of Eq. 4

Initially, we use Eq. 5 to derive the number of levels. We convert the derived number to the closest integer either by rounding up or down as necessary.

$$DvU = \log_{10}(total\_no\_of\_users\_in\_dataset) \quad (5)$$

Then, we use Eq. 6 to come with the number of co-rated items that will be assigned to the first level. The first level is the higher as shown in Eq. 8 and will be adjusted positively with a higher value.

$$DvI = \log_2(total\_no\_of\_items\_in\_dataset) \quad (6)$$

At the next step we divide the value from Eq. 6 with the value from Eq. 5 as shown in Eq. 7.

$$\frac{DvI}{DvU} \quad (7)$$

This gives the number of co-rated items that users should have to be allocated to a certain level. Additionally, if the number is not an integer it is converted to its closest. However, a threshold of 5 is used as the minimum number of co-rated items in order for a similarity value to be adjusted positively and an extra level is dynamically created: a user who does not have the minimum number of co-rated items (as derived from Eq. 7) is allocated to this extra level. At this level the similarity value is adjusted in a negative way. After the number of levels and the number of minimum and maximum co-rated is derived, then Eq. 8 is used to adjust the similarity values between users. Eq. 8 uses parts from Eq. 4 with modifications in order to adjust the similarities either in a positive or in a negative way.

$$Sim_{a,b}^{Proposed} = \begin{cases} Sim_{a,b}^{PCC} + Sim_{a,b}^{PCC}, & if \geq \frac{|Ia \cap Ib|}{M} \\ Sim_{a,b}^{PCC} + (\frac{Sim_{a,b}^{PCC}}{2}), & if \frac{|Ia \cap Ib|}{M} < m1 \text{ and } \frac{|Ia \cap Ib|}{M} \geq m2 \\ Sim_{a,b}^{PCC} + (\frac{Sim_{a,b}^{PCC}}{3}), & if \frac{|Ia \cap Ib|}{M} < m2 \text{ and } \frac{|Ia \cap Ib|}{M} \geq m3 \\ Sim_{a,b}^{PCC} + (\frac{Sim_{a,b}^{PCC}}{4}), & if \frac{|Ia \cap Ib|}{M} < m3 \text{ and } \frac{|Ia \cap Ib|}{M} \geq m4 \\ \ldots \\ \ldots \\ \ldots \\ Sim_{a,b}^{PCC} + (\frac{Sim_{a,b}^{PCC}}{n}), & if \frac{|Ia \cap Ib|}{M} < mn1 \text{ and } \frac{|Ia \cap Ib|}{M} \geq mn2 \\ Sim_{a,b}^{PCC} * (\frac{1 + (Sim_{a,b}^{PCC} * Sim_{a,b}^{PCC})}{n} - 1), & otherwise \end{cases} \quad (8)$$

### 3.1.1 Application example

In this section we deliver an example with real numbers in order to make the application of the dynamic multi-level method easier to understand. For this example, we use the data from the MovieTweetings dataset, with a further explanation of which can be found in section 4. This dataset has 39,363 users and 22,610 items.

The steps required to derive the levels are as follows:

1) Application of Eq. 5 to 39,363 gives DvU = 4.59
2) Rounding up of 4.59 gives 5
3) Application of Eq. 6 to 22,610 gives DvI = 14.46
4) Rounding down of 14.46 gives 14
5) Application of Eq. 7 gives $\frac{DvI}{DvU} = 14/5 = 2.8$
6) Rounding up of 2.8 gives 3

Therefore, the number of levels that adjust the similarity in a positive way is 5 (although 4 levels are actually used since the fifth level drops below the threshold of 5 co-rated items). In each level the number of co-rated items is 3 and the threshold of the minimum co-rated items is 5. Moreover, an extra level is added that adjusts the similarity in a negative way. Algorithm 1 gives an explanation of the levels created using the method.

**Algorithm 1:** Multiple levels derived from the MovieTweetings dataset

```
1) If (n>=14) { /* Level 1, the Top/Higher level*/
2) return (PCC+PCC);
3) }
4) else if (n>=11 && n<=13){ /* Level 2 */
5) return (PCC)+(PCC/2);
6) }
7) else if (n>=8 && n<=10){ /* Level 3 */
8) return (PCC)+(PCC/3);
9) }
```

```
10) else if (n>=5 && n<=7){ /* Level 4 */
11)     return (PCC)+(PCC/4);
12) }
13) else { /* Level 5 (Extra level) */
14)     return (PCC*(1/1+(PCC)*(PCC)))/6;
15) }
```

It is shown that number 14 is assigned to the first level as shown in line 1 of the code, where if two users have 14 or more co-rated items, the similarity gets the highest possible positive adjustment. Going down to level 2 is shown that the number of co-rated items is either 11, 12 or 13 with that being 1- minus from the top level and the number of items in this level being 3. In level 3, again the number of co-rated items is 3 and in level 4 the same. Thus, we start we 14 in the top level and then the number of co-rated items is always 3 until we reach a number of 5 co-rated numbers and the algorithm is terminated there. Finally, an extra level (level 5) is assigned for negative adjustments.

## 4. Experimental evaluation

In this section we explain the experimental evaluation, which was conducted on an Intel i3 2.13 GHz with 4GBs of RAM running windows 10 and the Java programming language. Furthermore, the experiments are based on three real datasets, MovieLens [10], MovieTweetings [31] and Epinions [11].

The experimental evaluation contains accuracy comparisons based on the Mean Absolute Error (MAE), the Normalized Mean Absolute Error (NMAE), a 5-fold cross evaluation based on the Root Mean Square Error metric (RMSE), Precision, Recall, F1 and Hit rate comparisons [6,32]. In section 4.1 the characteristics of the datasets used are described, in section 4.2 alternative methods used for benchmarking are explained, in section 4.3 an explanation of the metrics used is delivered and in sections 4.4 and 4.5 can be found the settings used and the results obtained.

### 4.1 Datasets

We have used three real datasets to conduct our experiments. The datasets are MovieLens 1 million, MovieTweetings and Epinions.

- **MovieLens.** This is a real dataset obtained from the University of Minnesota which contains 4000 movies, 6000 users and 1,000,000 ratings. This is a publicly available dataset associated with the online movie recommendation of the university. The data are in the form of [userid] [itemid] [rating] and all the rating values are from 1 to 5.
- **MovieTweetings.** This is a real dataset crawled from twitter which contains 22610 movies, 39363 users and 431,780 ratings. The data are in the form of [userid] [itemid] [rating] and all the rating values are from 0 to 10. MovieTweetings is publicly available.
- **Epinions.** This is a real dataset crawled from Epinions.com website. It is a dataset for general, e-commerce, product recommendation and the data are in the form of [userid] [itemid] [rating] and all the rating values are from 1 to 5. The dataset has 664,824 ratings submitted from 49,290 users on 139,738 items. This dataset is publicly available.

We have used these datasets due the fact that are used in previous research, are from different domains and use different rating scales. The key statistics of the datasets are shown below:

| Dataset | Users | Items | Ratings | Sparsity | Rating range |
|---------|-------|-------|---------|----------|--------------|
| MovieLens | 6,000 | 4,000 | 1,000,000 | 0.958 | 1-5 |
| MovieTweetings | 39,363 | 22,610 | 431,780 | 0.999 | 0-10 |
| Epinions | 49,290 | 139,738 | 664,824 | 0.999 | 1-5 |

## 4.2 Comparisons
We have used the following recommendation methods in our comparisons.

1. **PCC.** This method is defined in Eq. 1 and is used to calculate the statistical correlation between two users based on common ratings in order to determine a similarity value from -1 to 1.
2. **WPCC.** This method extends PCC by adding a constraint on the number of co-rated items. If that number is above the threshold, then the similarity is enchased and otherwise PCC is used. Eq. 2 defines WPCC.
3. **SPCC.** This method is also based on PCC. In this method the similarity value returned for users with small co-rated items is weaker. SPCC is defined in Eq. 3.
4. **Multilevel.** This is a method described in [22] and it uses a static multilevel approach of 4 levels to provide recommendations based on collaborative filtering.
5. **PLUS.** This is a method that is described in [26]. It uses a similarity method such as PCC and adjusts the similarity values between users using a power law function. For this method we use PCC and then apply the power law function, as described in the first two steps of the research article.

In collaborative filtering although accuracy is an issue, most important is the quality of the Top-N (e.g. first 10 or first 20) recommendations which is measured using the Precision and Recall. Thus, we use the first three and more established methods in our accuracy MAE, NMAE and 5-fold cross RMSE comparisons as benchmarks. In our quality comparisons based on Precision, Recall and F1 measure all the methods have been used as benchmarks.

## 4.3 Measures
In order to measure the accuracy of the recommendations we have used MAE, which is a widely used metric for measuring accuracy [6,32]. MAE is defined in Eq. 9, where *pi* is the predicted rating and *ri* is the actual rating in the summation. MAE is used to measure the deviation between the predicted ratings and the actual ratings. Lower values are better. Additionally, we have use the normalized MAE (NMAE) which is a measure that is independent of the rating scale since its values are from 0 to 1, thus making the comparison between datasets easier. Eq. 10 defines NMAE [8,33]. In NMAE *rmax* is the maximum rating value available and *rmin* is the minimum rating value available in the rating scale (e.g. 1-5). Moreover, in NMAE lower values are better. Additionally, RMSE is defined in Eq. 11, where *pi* is the predicted rating and *ri* is the actual rating in the summation. In RMSE lower values are better.

$$MAE = \frac{1}{n} \sum_{i=1}^{n} |pi - ri| \quad (9)$$

$$NMAE = \frac{MAE}{rmax - rmin} \quad (10)$$

$$RMSE = \sqrt{\frac{1}{n} \sum_{i=1}^{n} (pi - ri)} \quad (11)$$

However, in information retrieval systems use metrics such as Precision and Recall than can be used in recommender systems in order to measure the quality of the recommendations. In these metrics a higher value is better. Eq. 12 defines Precision and Eq. 13 defines Recall. Moreover, Eq. 14 defines the F1 measure which combines Precision and Recall [32–35]. In Precision, Recall and F1 measures higher values are better. Eq. 15 defines hit rate, which is a classification measurement that reflects the proportion of users for which at least one recommendation can be made.

$$Precision = \frac{Correctly\ recommended\ items}{Total\ recommended\ items} \quad (12)$$

$$Recall = \frac{Correctly\ recommended\ items}{Relevant\ items} \quad (13)$$

$$F1 = 2 \cdot \frac{precision \cdot recall}{precision + recall} \quad (14)$$

$$hitrate_u = \begin{cases} 1 : if\ hits_u > 0 \\ 0 : else \end{cases} \quad (15)$$

### 4.4 Settings
For the experiments the following settings have been used.

- **WPCC.** For the MovieLens dataset this number has been set to 50, for the MovieTweetings dataset this has been set to 10 and for the Epinions dataset this value has been set to 5. (Several experiments with the WPCC method have shown that the values used provide more accurate recommendations when compared to alternatives. Unfortunately, WPCC in its current form only takes static values).
- **MAE.** For the evaluation based on MAE as shown in Fig. 1, Fig. 2 and Fig. 3 the datasets have been split into two parts. The first one is the training part consisting of 80% and the testing part consisting of the rest 20%. For the results in Fig. 4, Fig. 5 and Fig. 6, 60% has been used for training and 40% for testing.
- The **t** and **y** values of the proposed (static) method have been set to 10 and 0.20 respectively for the MovieLens and MovieTweetings datasets. For the Epinions dataset the values have been set to 5 and 0.15 respectively. (Several experiments with the t and y values have shown that these values provide more accurate recommendations when compared to alternatives. Unfortunately, this is a major issue when static methods are used).
- **PLUS.** This method is based on a function $f(x) = \alpha x^\beta$ where x is the similarity value between two users. Two tests took place with the value of α and β being 100 and 2 respectively in the first test, while in the second test the values 80 and 5 have been used, which provide lower quality results. (This is a static method and the set of numbers α = 100 and β = 2 provided results of higher quality when we run a number of tests in order to determine a set of number that provide results of high quality).

- **hit rate.** In order to measure hit rate, we used Eq. 15 to calculate the number of users that at least one recommendation can be provided, then divided this number by the total number of users in the dataset and multiplied by 100 to deliver a percentage range.

**4.5 Experimental Results**

The MAE results obtained from the MovieLens dataset are shown in Fig. 1. It is shown that our proposed method outperforms the alternatives and that as the user neighborhood grows then our method becomes more effective. In Fig. 2 the results from the MovieTweetings dataset are shown. It is shown that our proposed method outperforms all the alternatives in every user neighborhood size. In Fig. 3 the results from the Epinions dataset are shown. It is shown that our proposed method outperforms all the alternatives in every user neighborhood size, except the fisrt case. In all cases in the figures and tables k is the number of the nearest users forming the neighborhood.

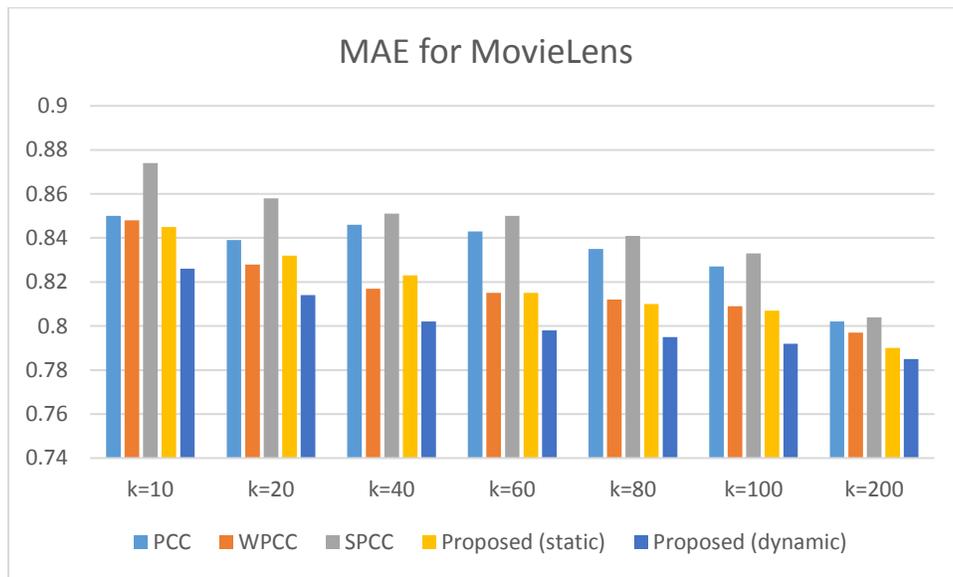

**Fig. 1** MAE results for MovieLens

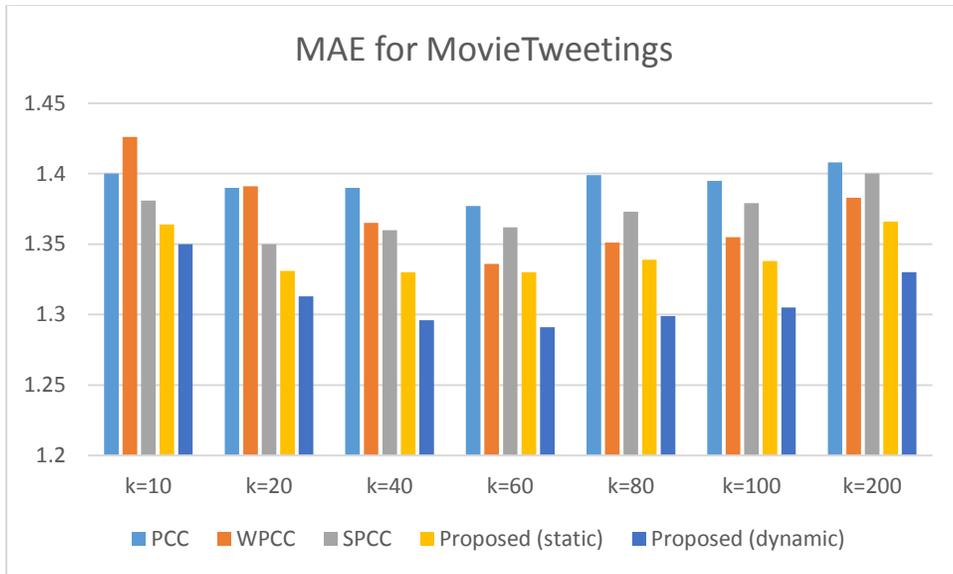

**Fig. 2** MAE results for MovieTweetings

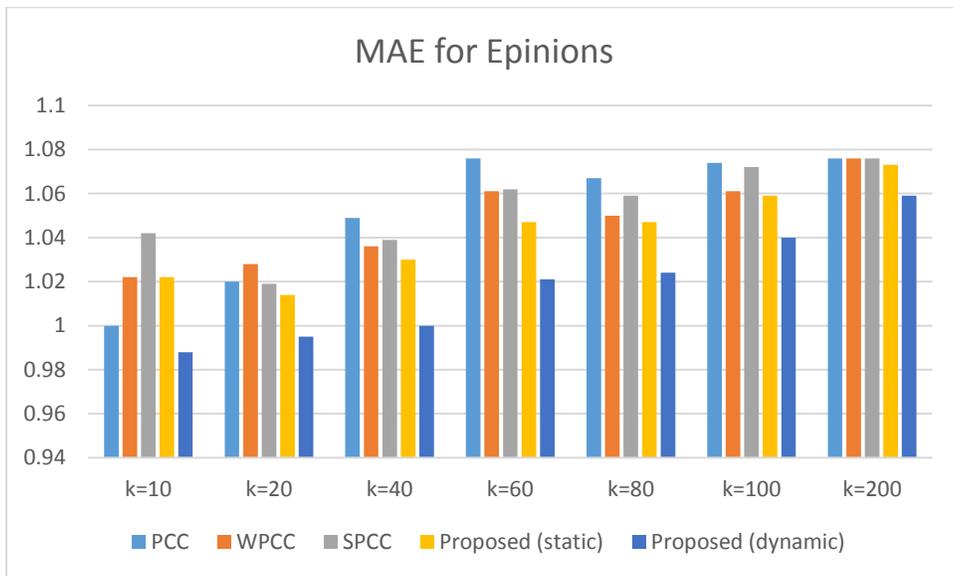

**Fig. 3** MAE results for Epinions

Furthermore, in figures 4, 5 and 6 we can see a comparison between our proposed method and PCC that is based on a 60% training set and a 40% test set, for MovieLens, MovieTweetings and Epinions.

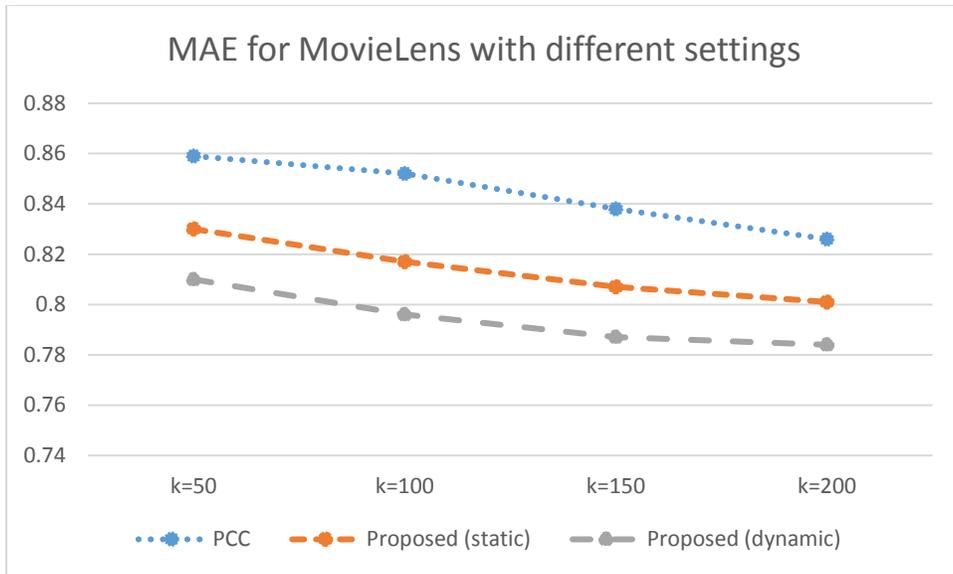

**Fig. 4** MAE results for MovieLens with different settings

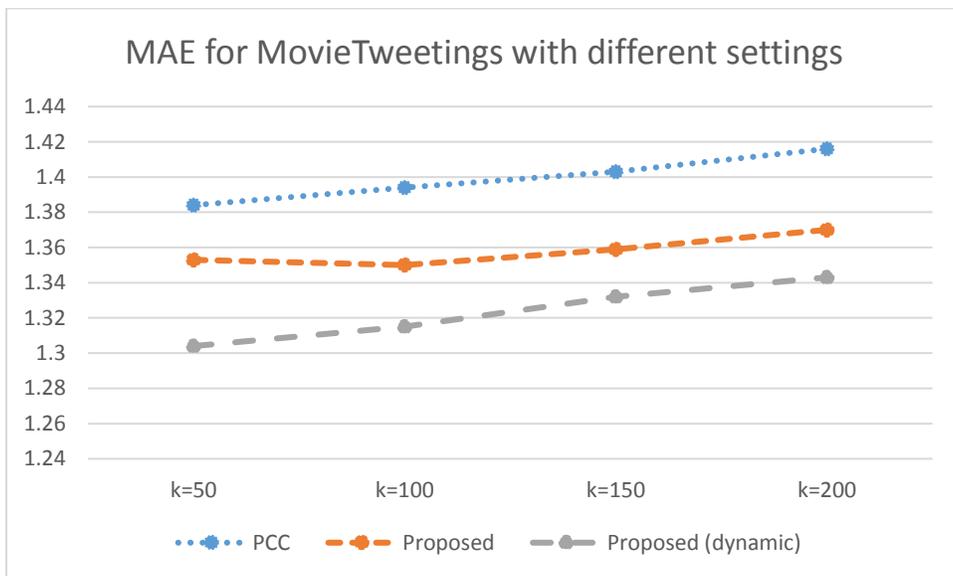

**Fig. 5** MAE results for MovieTweetings with different settings

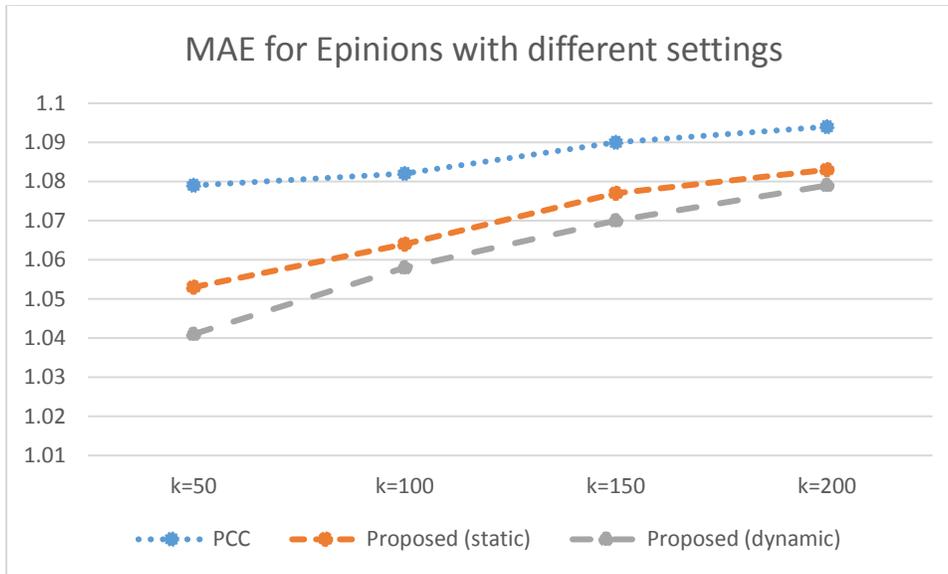

**Fig. 6** MAE results for Epinions with different settings

Additionnally in tables 2 and 3 we can see a comparison using the NMAE measure, which makes comparison between different ratings scales, and therefore datasets, easier. In table 2, k is set to 40 and in table 3 to 80. The dataset is split to 80% for training and 20% for testing.

| **NMAE Comparisons** *k=40* | | | |
|---|---|---|---|
| | MovieLens | MovieTweetings | Epinions |
| PCC | 0.2215 | 0.139 | 0.2622 |
| WPCC | 0.2042 | 0.1365 | 0.2590 |
| SPCC | 0.2127 | 0.1360 | 0.2597 |
| Proposed (static) | 0.2057 | 0.1330 | 0.2575 |
| Proposed (dynamic) | 0.2005 | 0.1296 | 0.2500 |

**Table 2** NMAE comparison with k=40

| **NMAE Comparisons** *k=80* | | | |
|---|---|---|---|
| | MovieLens | MovieTweetings | Epinions |
| PCC | 0.2087 | 0.1399 | 0.2667 |
| WPCC | 0.2030 | 0.1351 | 0.2625 |
| SPCC | 0.2102 | 0.1373 | 0.2647 |
| Proposed (static) | 0.2025 | 0.1339 | 0.2617 |
| Proposed (dynamic) | 0.1987 | 0.1299 | 0.2560 |

**Table 3** NMAE comparison with k=80

Table 4 lists the results obtained from the 5-fold cross validation tests based on the RMSE metric and with comparisons to the traditional methods. In this experiment the datasets have been split to 5 nearly equal sized parts where each part is used as a traning for the next part which is the testing part. This takes places until we reach 5 tests.

|  | 5-fold cross validation based on RMSE | | | | | |
|---|---|---|---|---|---|---|
| Dataset | 5 fold cross-validation | PCC | WPCC | SPCC | Proposed (static) | Proposed (dynamic) |
| MovieLens | 1 | 1.146 | 1.102 | 1.139 | 1.102 | 1.092 |
|  | 2 | 1.110 | 1.095 | 1.134 | 1.077 | 1.075 |
|  | 3 | 1.130 | 1.095 | 1.128 | 1.071 | 1.093 |
|  | 4 | 1.144 | 1.099 | 1.147 | 1.078 | 1.083 |
|  | 5 | 1.105 | 1.084 | 1.144 | 1.076 | 1.084 |
|  | Average | 1.127 | 1.095 | 1.138 | 1.081 | 1.085 |
| MovieTweetings | 1 | 1.907 | 1.898 | 1.849 | 1.844 | 1.800 |
|  | 2 | 1.908 | 1.897 | 1.850 | 1.818 | 1.773 |
|  | 3 | 1.858 | 1.855 | 1.825 | 1.777 | 1.777 |
|  | 4 | 1.869 | 1.858 | 1.850 | 1.846 | 1.778 |
|  | 5 | 1.874 | 1.867 | 1.848 | 1.827 | 1.752 |
|  | Average | 1.883 | 1.875 | 1.844 | 1.822 | 1.776 |
| Epinions | 1 | 1.373 | 1.316 | 1.370 | 1.337 | 1.261 |
|  | 2 | 1.384 | 1.306 | 1.316 | 1.279 | 1.273 |
|  | 3 | 1.362 | 1.303 | 1.397 | 1.364 | 1.299 |
|  | 4 | 1.380 | 1.312 | 1.382 | 1.357 | 1.270 |
|  | 5 | 1.358 | 1.292 | 1.352 | 1.328 | 1.273 |
|  | Average | 1.372 | 1.306 | 1.363 | 1.333 | 1.275 |

**Table 4** RMSE 5-fold cross validation results

Table 5 presents a comparison of our proposed method and alternatives. The comparisons are based on Precision, Recall and F1. The number of nearest neighbors, $k$, is set to 150 and the number of the requested recommendations $r$ is set to 20.

| Precision, Recall and F1 comparisons of the proposed static and dynamic methods with alternatives $k=150, r=20$ | | | | | | | | | |
|---|---|---|---|---|---|---|---|---|---|
|  | MovieLens | | | MovieTweetings | | | Epinions | | |
|  | Precision | Recall | F1 | Precision | Recall | F1 | Precision | Recall | F1 |
| PCC | 0.106 | 0.124 | 0.114 | 0.101 | 0.190 | 0.130 | 0.032 | 0.041 | 0.036 |
| WPCC | 0.142 | 0.160 | 0.145 | 0.114 | 0.209 | 0.142 | 0.038 | 0.048 | 0.042 |
| SPCC | 0.102 | 0.120 | 0.108 | 0.103 | 0.195 | 0.134 | 0.031 | 0.041 | 0.035 |
| Multilevel | 0.180 | 0.209 | 0.193 | 0.126 | 0.235 | 0.160 | 0.039 | 0.049 | 0.043 |
| PLUS α=100 β=2 | 0.174 | 0.202 | 0.187 | 0.120 | 0.225 | 0.157 | 0.038 | 0.050 | 0.043 |
| PLUS α=80 β=5 | 0.153 | 0.179 | 0.165 | 0.117 | 0.220 | 0.153 | 0.037 | 0.048 | 0.041 |
| Proposed (static) | 0.143 | 0.167 | 0.148 | 0.116 | 0.217 | 0.150 | 0.037 | 0.048 | 0.041 |
| Proposed (dynamic) | 0.173 | 0.202 | 0.181 | 0.127 | 0.237 | 0.164 | 0.043 | 0.056 | 0.048 |

**Table 5** Precision, Recall and F1 comparisons

Table 6 presents the hit rate comparison between our proposed method and alternatives. The number of nearest neighbors, *k,* is set to 150 and the number of the requested recommendations *r* is set to 20. Furthermore, at least one recommendation needs to be provided to a user in order to have a hit rate value greater than zero for this particular user. Then, the output represents the overall value between all users.

| **Hit rate comparisons of the proposed static and dynamic methods with alternatives** <br> *k=150, r=20* | | | | | | | | |
|---|---|---|---|---|---|---|---|---|
| | PCC | WPCC | SPCC | Multilevel | PLUS $\alpha=100$ $\beta=2$ | PLUS $\alpha=80$ $\beta=5$ | Proposed (static) | Proposed (dynamic) |
| MovieLens | 50% | 52% | 50% | 57% | 56% | 55% | 55% | 57% |
| MovieTweetings | 4.6% | 4.7% | 4.7% | 4.9% | 4.9% | 4.8% | 4.8% | 4.9% |
| Epinions | 1.2% | 1.3% | 1.2% | 1.2% | 1.3% | 1.3% | 1.3% | 1.4% |

**Table 6** Hit rate comparisons

## 5. Discussion

Collaborative filtering is widely used in e-commerce to provide suggestions of items or services to users, which include the recommendations of movies, books, general products or users to users. Therefore, users can benefit from finding relevant items without the burden of manually searching and companies could grow extra profit from higher sales of items. Additionally, the computational needs for the company are reduced, since the users normally will not bother making several search attempts to find relevant items. As a result, recommender systems and collaborative filtering are valuable tools for both a user and a company and if an algorithm can produce quality results is beneficial for both sides. Thus, we propose a method that is based on the characteristics of the dataset and adds constraints that improve the quality of the Top-N recommendations. Our method is based on constraints that can be added either manually or dynamically to the settings of the PCC recommendation method. In the case of manually addition of constraints, a developer should perform several tests in order to find the optimal settings for the particular data. On the other hand, in the case of the dynamic method this is not necessary, since a series of steps takes place in order to do this.

In our experimental accuracy MAE results, it is shown that the proposed (static) method marginally outperforms the other recommendation methods, although this is not the case in every test and dataset. Instead, the proposed (dynamic) method performs better, since it outperforms the alternative recommendation methods under all neighborhood sizes, datasets and when different settings are used. Figures 1, 2 and 3 represent the accuracy results for the MovieLens, the MovieTweetings and Epinions dataset respectively and include the proposed methods along with comparisons to alternatives. Lower values represent more accurate results in the recommendations provided, thus the closest the value is to zero better recommendations are shown to the user. Figures 4, 5 and 6 show accuracy results that are based on different settings in order to support the effectiveness of our proposed method. Still, under these different settings our method performs well. However, it should be noted that different ratings scales (e.g. 0-10, 1-5 etc.) produce different MAE accuracy numerical scales, thus making the comparison across different datasets harder to interpret. The NMAE metric solves this problem by converting all MAE values in a zero to one scale and its outputs can be used to compare the gains or losses from different recommendation methods across different datasets. Therefore, tables 2 and 3 in the experimental evaluation can assist by showing comparisons based on MAE and all datasets and therefore making the comparison of a method in different datasets easier. Our NMAE results are based on converted and previously explained MAE results to show that our method produces similar improvement when applied in different datasets. Furthermore, the results of the 5-fold cross validation based on RMSE are shown in table 4 and validate our proposed method, since in all tests our proposed method outperforms all the traditional recommendation methods.

Although accuracy is considered an important aspect when evaluating recommender systems, alternative evaluation methods can supplement these results to support an approach. These include the Precision, Recall and F1 metrics that have been described in section 4. Table 6 is a comparison of our static and dynamic methods with alternative recommendations methods. It is shown that our dynamic method produces similar results to the state of the art.

The experimental results show that the proposed (dynamic) method performs well both in terms of accuracy and quality of the Top-N recommendations. In addition, it produces better results when compared to alternative recommendation methods. Furthermore, it can be utilized by online companies to reduce the load in their servers and assist users in decision making by providing more accurate recommendations in different domains. Also, someone could choose to use the proposed (static) method that marginally produces better results from the alternatives, due to it is less complex. However, this method does not produce optimal results in every case. The proposed (dynamic) method, on the other hand, produces stronger results and fits well in expert decision making in online environments that utilize collaborative filtering.

## 6. Conclusions and future work

Recommender systems and collaborative filtering have emerged as a standalone research area, with their use found in several online environments. Collaborative filtering recommender systems are being used in the web to provide recommendations of items, services or even users to users. However, accuracy problems exist when predictions for recommendations are being made. People utilize recommender systems when online to assist them in the decision making process by substituting the opinion of human experts. The recommendations delivered by a recommender system need to be as accurate as possible, since the job of recommendation systems is to substitute a human expert. Thus, we have proposed a recommendation method that improves the quality of the requested collaborative filtering recommendations. Our proposed method is based on constraints that accompany the classical PCC method. By adding constraints and make adjustments of positive and negative nature to user similarity values the accuracy of the recommendations is improved. The main accomplishment of our method is that it is not necessary to rely on extra or additional information and can utilize information found in the database of ratings, such as the number of items and users in order to improve the accuracy and the output quality of the recommendations.

In addition to what has been discussed we would like to further improve our work in the future by enchasing the robustness of the system against shilling attacks by utilizing a method that can prevent attacks such as the injection of fake profiles that are used to promote certain items. Moreover, with the increased popularity of mobile technologies and location-based services (LBS), collaborative filtering can be used as a part of such systems to provide recommendations in mobile environments. A future work towards this direction will include the proposal of a method that can be used effectively in mobile devices. Furthermore, we aim to focus on a system that can preserve user privacy produce recommendations of high accuracy and quality and can be applied both in web and mobile environments.